\documentclass[10pt,letterpaper]{article}
\usepackage[top=0.85in,left=2.75in,footskip=0.75in]{geometry}

\usepackage{amsmath,amssymb}

\usepackage{changepage}

\usepackage{textcomp,marvosym}

\usepackage{cite}

\usepackage{nameref,hyperref}

\usepackage[right]{lineno}

\usepackage[nopatch=eqnum]{microtype}
\DisableLigatures[f]{encoding = *, family = * }

\usepackage[table]{xcolor}

\usepackage{array}

\newcolumntype{+}{!{\vrule width 2pt}}

\newlength\savedwidth

\raggedright
\usepackage[aboveskip=1pt,labelfont=bf,labelsep=period,justification=raggedright,singlelinecheck=off]{caption}

\makeatletter
\renewcommand{\@biblabel}[1]{\quad#1.}
\makeatother

\usepackage{lastpage,fancyhdr,graphicx}
\usepackage{epstopdf}
\fancyheadoffset[L]{2.25in}
\fancyfootoffset[L]{2.25in}
\begin{document}
\vspace*{0.2in}

\begin{flushleft}
{\Large
\textbf\newline{A low-cost, open-source maskless photolithography stepper for microfabrication} 
}
\newline
\\
B. Joel Gonzalez\textsuperscript{*},
Elio Bourcart,
J. Kent Wirant,
Rachel Jiang,
Peiyu Lin,
Michael Casey Honaker,
Toby Yang,
Jessica Alacce,
Jocelyn Xu,
Michael Juan,
Justin Wang,
William Gao,
Matthew T. Moneck
with the Hacker Fab\textsuperscript{\textpilcrow}
\\
\bigskip
Department of Electrical and Computer Engineering, Carnegie Mellon University, Pittsburgh, PA, USA
\bigskip

%
%

\textpilcrow Membership list can be found in the Acknowledgments section.

* Corresponding author: bgonzale@andrew.cmu.edu

\end{flushleft}
\section*{Abstract}
Photolithography is a key part of modern semiconductor process flows, and photolithography steppers have been used for decades to achieve precise patterning for device fabrication. However, these tools are often large and expensive, which restricts their use to industry and well-funded university laboratories. In this paper, we propose a \$3000 maskless photolithography stepper that is affordable, open-source, and easy to assemble. The stepper, which uses a Digital Light Processing (DLP) projector as its optical engine, is able to achieve an optical resolution of under 2 microns. The stepper also features a motorized nanopositioning system, which is able to align features with sub-micron precision. RF-DETR, an object detection computer vision model, is used to detect fiducials on chip for automatic fine-grain alignment and stitching of patterns. These capabilities allow for use of the stepper for microfabrication to produce micron-scale devices such as NMOS transistors.


\section*{Introduction}
Lithography is a critical step in any microfabrication process, which determines the scale of the device and the achievable circuit complexity. Photolithography, which uses light to pattern features onto a substrate, has been used to produce transistors since the mid-20th century \cite{lahtrop1947}, and the photolithography stepper has been used in semiconductor fabrication since the 1970s \cite{stepper1975}. The stepper is a tool that performs a ``step and repeat" action, covering an entire chip or wafer one exposure at a time until it has patterned the entire surface of the substrate. Such tools revolutionized the semiconductor industry, allowing for higher throughput of fabricated devices, thus paving the way for the development of the micro- and nanoelectronics of today.

However, the cost of these tools creates a large barrier to entry to microfabrication. Modern commercial instruments such as the Heidelberg MLA150 cost several hundred thousands of dollars, restricting their use to academic or industrial settings \cite{mla2025}. Nanofabrication facilities such as the Bertucci Nanofabrication Laboratory at Carnegie Mellon University contain several photolithography systems that also require several thousands of dollars per year in upkeep and maintenance costs \cite{nanofab2025}. Used products can cost upward of \$50,000 \cite{labx2025}. For smaller research laboratories or groups that require photolithography tools for prototyping their own devices, the current state of the art technology is prohibitively expensive.

Prior works across various research groups have partially addressed this accessibility problem via “DIY” approaches, which includes either augmentation of existing systems or building systems from scratch. Contact lithography systems achieving 1-2 $\mu$m patterning resolution have been produced with materials costing around \$12,000 from scratch \cite{https://doi.org/10.1002/adem.202401228}, but these still require photomasks, which themselves are expensive to produce and have fixed designs. Maskless designs employing LCDs or commercially available LCD projectors for patterning have existed since at least the early 2000s \cite{10.1117/1.1531976}, with from-scratch builds demonstrating patterning resolutions of 14 $\mu$m for ¥500,000 in 2015 \cite{HORIUCHI201537}. Other maskless implementations used projection systems with a digital micromirror device (DMD) to achieve patterning resolutions down to 2 $\mu$m for a material cost of \$3,000 \cite{mi15010039}. However, these designs are still not truly accessible since they may be difficult to reproduce without freely available design documentation and accompanying software.

Open-source hardware provides a means to solve this problem. In other fields of engineering, open-source hardware has made a massive impact on the ability to conduct research at a reduced cost \cite{optics2013} \cite{syringe2014} \cite{reprap2025}. Even within the field of device fabrication, there has been research into open-source hardware for the deposition of materials (specifically chemical vapor deposition) \cite{cvd2019}. Chip design and fabrication are met with to open-source or democratization projects such as FOSSi \cite{fossi2025} and Tiny Tapeout \cite{TT2025}, but microfabrication hardware often has limited access. The Hacker Fab, an initiative born at Carnegie Mellon University in 2023, seeks to develop open-source hardware for chip fabrication \cite{hfab2025}.

In this work, we present the first fully open-source maskless photolithography stepper for device microfabrication made by the Hacker Fab, with patterning resolution in the $2 \mu m$ range, a low material cost of \$3,000, and with fully automatic single-layer alignment and exposure, including stitching of multi-exposure designs. To our knowledge, there does not exist another open-source work with these metrics, which are comparable to state-of-the-art in-house designs and commercially available clean-room tools. The maskless capability of the stepper allows for rapid prototyping of devices compared to using expensive and time-consuming photomasks, at the expense of smaller exposures requiring a stitching software algorithm. The performance and capabilities of the Hacker Fab stepper are assessed and compared to commercially available tools found in a clean-room setting. The Hacker Fab stepper consists of hardware components that can be easily acquired from online vendors, with the Bill of Materials and assembly instructions available online for free on the Hacker Fab GitBook page \cite{gitbook2025}. The current iteration of the stepper is shown in Fig. 1. In pushing open-source hardware, the barrier to entry can be lowered so that anyone with access to the Internet can learn and build their own fabrication tools for semiconductor devices.

\begin{figure}[h!]
    \centering
    \includegraphics[width=1.0\linewidth]{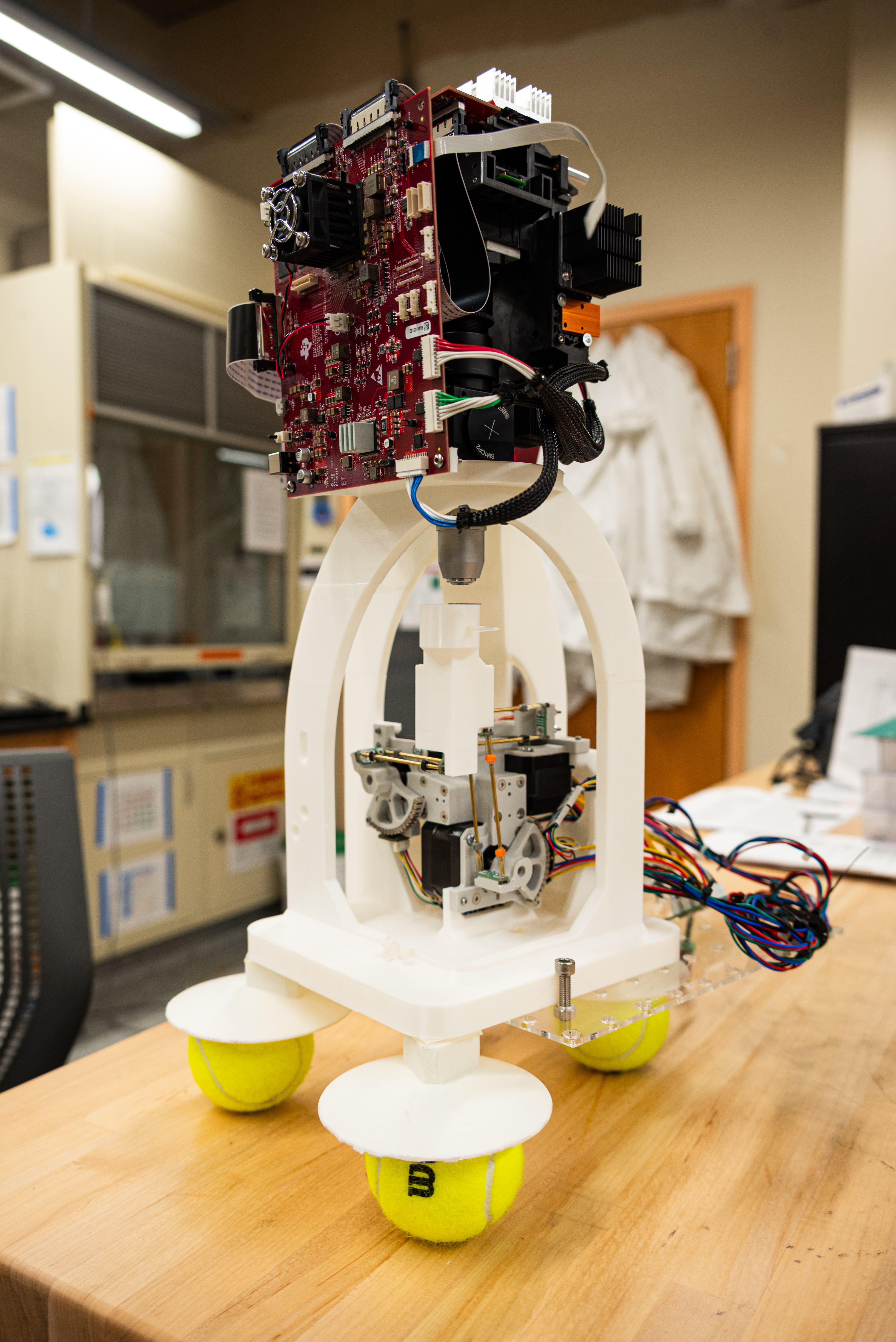}
    \caption{\textbf{The Hacker Fab Stepper.}}
    \label{fig:enter-label}
\end{figure}

\newpage
\section*{Materials and Methods}

The Hacker Fab stepper consists of several subsystems (each of which is described in further detail below), which are integrated into a fully functional photolithography tool for about \$3,000. The Bill of Materials and most up-to-date build instructions can be found on the GitBook page \cite{gitbook2025} and the code can be found on GitHub \cite{github2025}.

An overview of the stepper is shown in Fig. 2. The system consists of optical, electromechanical, and software subsystems. The optical engine is responsible for generating a near-UV light source that can then be focused onto a chip for patterning. The electromechanical components of the stepper consist of a nanopositioner system actuated by stepper motors driven by a microcontroller, which allows users to precisely position their chip for exposures. Lastly, the software is a set of Python scripts with a user interface that is run on a personal computer connected to both the evaluation board and the microcontroller. Each of these systems is discussed in further detail in the following subsections.

\begin{figure}[h!]
    \centering
    \includegraphics[width=1.0\linewidth]{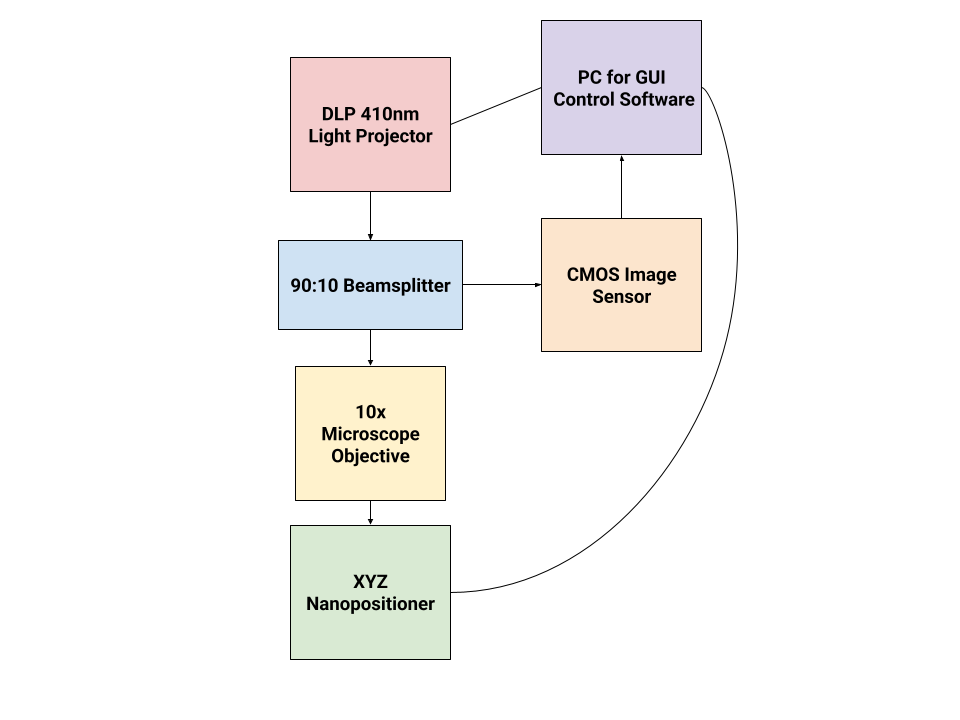}
    \caption{\textbf{Stepper System Overview.} Diagram showing the various optical, electromechanical, and software components of the lithography system.}
    \label{fig:enter-label}
\end{figure}

\subsection*{Optical Stack}

The optical stack of the stepper, shown in Fig. 3, consists of a near-UV LED light source, a Texas Instruments DLP projector, a camera sensor, and miscellaneous optical components. The DLP projector model is the DLPDLCR471TPEVM (or 471TP for short), containing a DLP471TP 4K digital micromirror device (DMD), DLPC6540 HSSI display controller, and DLPA3005 PMIC/LED driver \cite{471TP}; together, these act as a reconfigurable photomask with a resolution of 1920 by 1080 pixels. When combined with the evaluation kit's optical engine and HDMI interface, users may project an arbitrary image from their computer for custom patterning without the need for a photomask. A pattern of near-UV light is projected from this system and fed into a tube lens.

\begin{figure}[h!]
    \centering
    \includegraphics[width=1.0\linewidth]{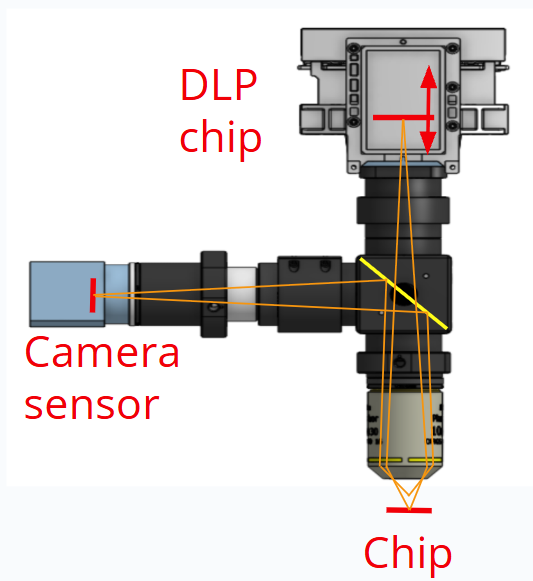}
    \caption{\textbf{Optics Diagram.} Ray diagram of optics stack on the current stepper model.}
    \label{fig:enter-label}
\end{figure}

This tube lens connects to a 90:10 beam splitter, which is oriented such that it reflects from the surface of the chip to the camera. The beam splitter is then connected to two additional tube lenses. One of these tube lenses connects to a C-mount Basler Ace camera, which is used to observe the area being patterned on the chip. A modular design enables support for other cameras as well, such as a FLIR camera or an AmScope camera. The other tube lens is connected to a 10x magnification objective with a 0.25 numerical aperture, which is used to magnify the image being projected onto the chip. The image projected onto the chip's surface is approximately 1.04mm by 0.58mm.

The light source itself is a 410nm Lumiled near-UV LED, which is soldered onto a copper-core PCB as shown in Fig. 4 \cite{digikey2025}. The blue LED channel PCB in the projector is swapped for this near-UV LED PCB, to allow for greater sensitivity to the wavelengths of light required for exposure of our AZ P4000 series photoresist. The PCB itself allows for the placement of up to four near-UV LEDs, although during testing, it was determined that only two LEDs in series were necessary for successful patterning. The components of the optics stack were chosen to be compatible with near-UV light to reduce the effect of light being absorbed by the glass components; if needed, the time of exposure can be adjusted to get the desired photolithographic results.

\begin{figure}[h!]
    \centering
    \includegraphics[width=1.0\linewidth]{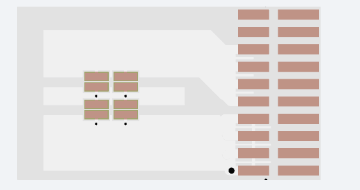}
    \caption{\textbf{Near-UV LED PCB.} The board allows for up to four near-UV LEDs to be placed.}
    \label{fig:enter-label}
\end{figure}

As of the writing of this paper, the 471TP evaluation kit has frequently been out of stock and has long lead times. One possible solution is to use an alternative projector, such as the DLP4710EVM-LC, as the light engine \cite{DLP4710}. Though this requires some modifications to the optics stack (details can be found on our GitBook page \cite{gitbook2025}), and is slightly more expensive, this presents an alternative route for building the stepper system. Future considerations for the design of the optical stack are provided in the Discussion section of this paper.

\subsection*{Electromechanical Design}

The electromechanical design of the stepper is based on the Open Micro-Manipulator project with several improvements \cite{nanopositioner2026}, shown in Fig. 5. The main mechanical structure is a delta robot driven directly by three NEMA 17 stepper motors. This parallel arrangement is mechanically stiff and is built from a small number of low-cost parts. The three motor shafts are connected to the end effector by six linkage rods cut from 2 mm solid brass rod. Each rod is sanded to a common length in a jig, and both ends are turned down to a 1 mm tip into which a shallow spherical seat is pressed, so that all six rods share a matched effective length. The joints themselves are 1 mm steel balls mounted on ball-joint plates on each motor crank and on the end effector; the dented rod tips seat directly on these balls and are held in preload by stretch cord hooked over 3D-printed collets on each rod. This arrangement eliminates the backlash of conventional rod-end bearings while keeping joint friction low.

\begin{figure}[h!]
    \centering
    \includegraphics[width=1.0\linewidth]{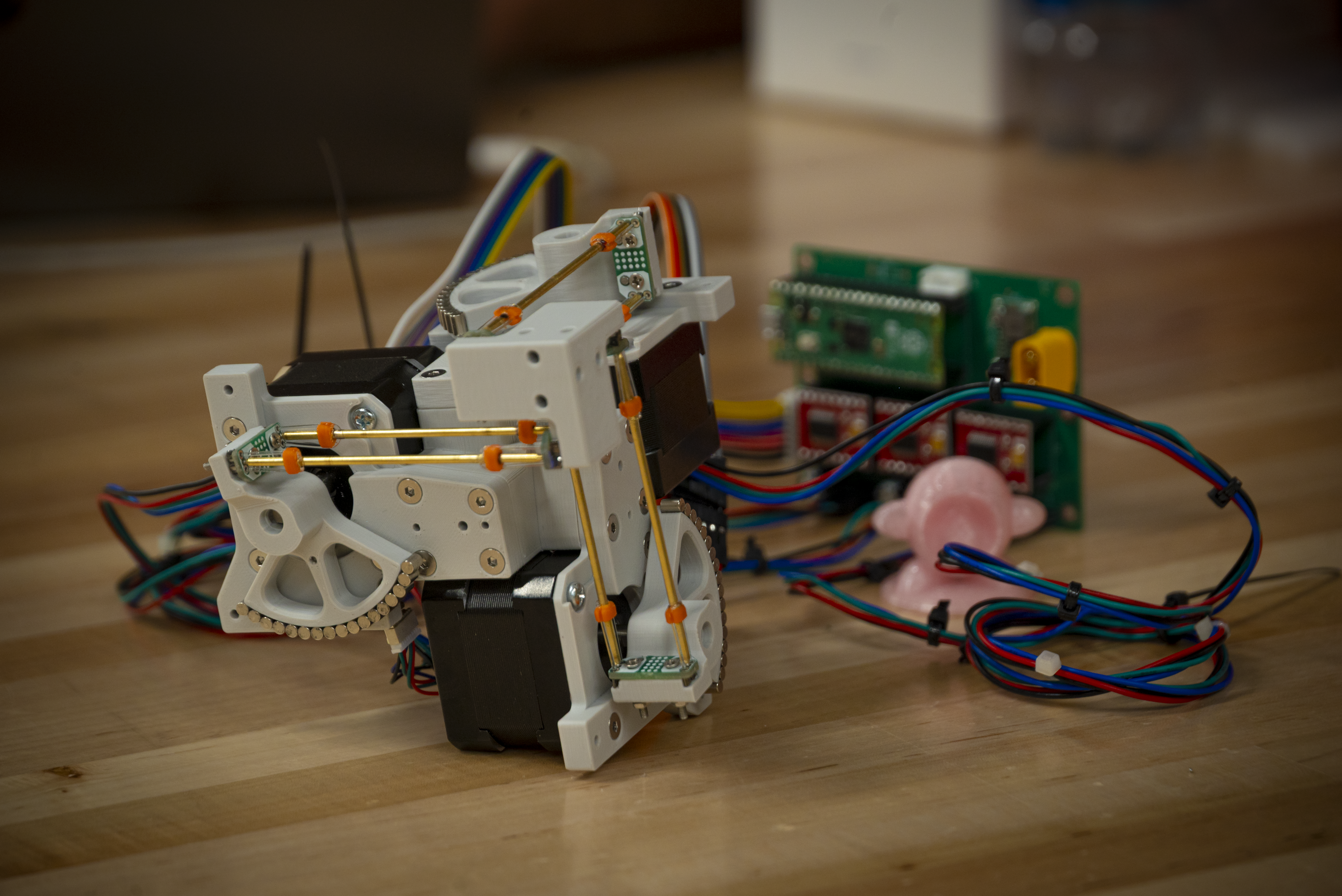}
    \caption{\textbf{XYZ nanopositioner.} Isolated shot of the nanopositioner, showing stepper motor assembly in the foreground and electronics control board in the background.}
    \label{fig:enter-label}
\end{figure}

Motion control runs on a Raspberry Pi Pico 2 (RP2350); however, any sufficiently strong MCU should suffice. Each two-phase stepper is driven by a dual H-bridge, with one bridge per motor phase, so the firmware sets both phase currents directly rather than issuing step/direction pulses to a conventional stepper driver. Each motor crank sweeps only about 90° between mechanical end stops, and this limited stroke defines the reachable workspace of the delta linkage. Rotor position is measured by MT6835 21-bit magnetic AMR encoders read over a shared 8 MHz SPI bus. Rather than reading a single magnet on the motor shaft, each motor has an arc of 17 alternating magnets spanning between the mechanical endstops. This array means the encoder completes one full electrical period every two magnets, thus a single period spans only 12\% of total crank rotation. The trade-off is that the encoder reading repeats roughly every eight periods across the stroke so it is not absolute on its own. Absolute position is instead established at startup by homing against the end stop. A steel dowel pin on each crank travels in a groove in the motor mount, and during homing each joint is driven at reduced current until the pin reaches the end of its groove, giving a repeatable mechanical reference. Small irregularities in the stepper make the encoder to movement angle mapping nonlinear, so each joint is calibrated once by sweeping it through 92\% (i.e. about 82 degrees of crank movement) of its full travel from the home position and storing a movement correction lookup table in flash.

To achieve motion resolution finer than the physical step size, the firmware treats each stepper as a two-phase synchronous motor and holds it at arbitrary positions between full steps (sub-stepping). The two phase currents are set to the sine and cosine of a commanded electrical angle using a 20 kHz PWM signal with 12-bit duty-cycle resolution. The PWM quantization therefore sets the theoretical resolution limit as the encoders read at significantly higher resolution. With the phase amplitude limited to 80\% of full scale due to heating concerns, the sine and cosine duty cycles are drawn from 3276 discrete levels, which divides one electrical revolution into approximately 20,600 distinguishable field angles. For a 1.8° motor with 50 pole pairs, this corresponds to a mechanical sub-step of 3.5E-4 degrees, or roughly 5,150 sub-steps per full step. Our delta geometry uses a 13 mm motor crank radius, so this angular sub-step maps to a theoretical linear resolution at the end effector of approximately 79 nm. This figure describes theoretical resolution and does not account for the holding force of the stepper magnets, nonlinearities in the motor response, or finite precision in the sine/cosine lookup tables used to calculate the phase currents.

\subsection*{Movement Validation}

Positional accuracy was measured optically and in situ through the lithography objective, so that the recorded quantity is the relative motion between the chip and the optical column rather than the position reported by the stage. A patterned chip mounted on the end effector was imaged at each stop, with a pixel scale of 0.62 $\mu$m per pixel calibrated against a USAF resolution target. Displacement between two stops was estimated by matching SIFT features between images and fitting a partial affine transform (translation, rotation and uniform scale) under RANSAC. An example of the feature detection and correlation can be seen in Fig. 6. Three images were captured at each stop and compared pairwise against the three images at the next stop, and each reported displacement is the mean of these nine comparisons. This averaging reduces random registration error and the effect of vibration at the moment of capture, although the nine comparisons share images and are not statistically independent.

\begin{figure}[h!]
    \centering
    \includegraphics[width=1.0\linewidth]{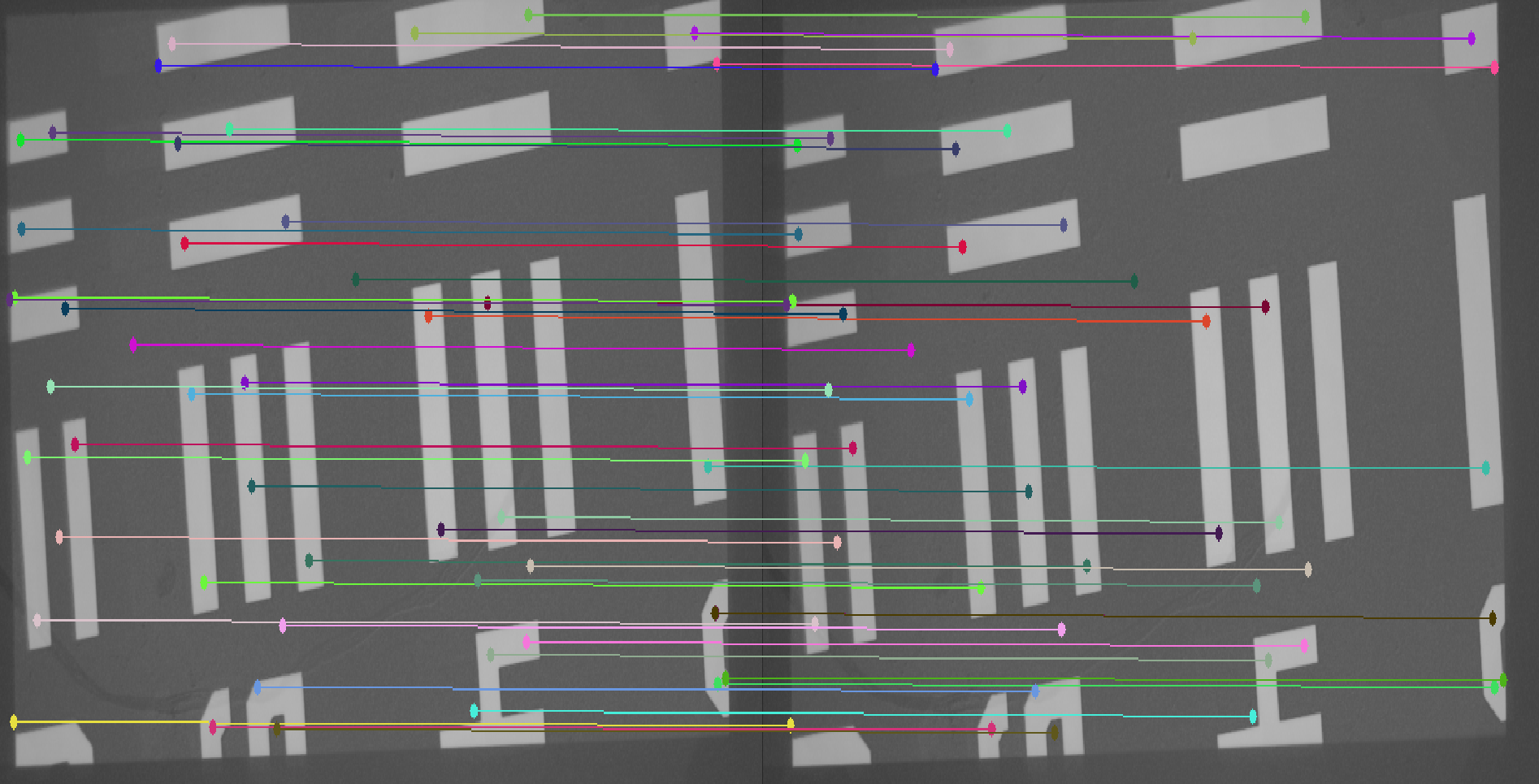}
    \caption{\textbf{Movement validation.} Image capture of displacement measurement using SIFT on a USAF resolution target.}
    \label{fig:enter-label}
\end{figure}

The test comprised 652 moves: 600 moves at commanded distances of 1, 10, 50, 100 and 250 $\mu$m, and 52 interleaved zero-length control moves used to estimate the noise of the measurement itself. At each distance, the stage was moved along X, Y and a 45 degree diagonal in a four-move cycle (two moves out and two moves back), giving 12 distinct moves that were each repeated twice at each of five feed rates between 0.1 and 25 mm/s. Cross-coupling from Z motion was tested separately with 10 cycles of 50 $\mu$m Z excursion and return, with the X-Y setpoint held constant. One issue with Z validation is that a 50 $\mu$m z move defocuses the sample, preventing reliable feature extraction thus displacement was evaluated only across complete down-and-up cycles. Vibration was additionally examined by optical-flow analysis of video recorded through the same optics, but the camera's maximum frame rate of approximately 140 Hz restricts this analysis to frequencies below roughly 70 Hz.

Zero-length control moves produced an apparent displacement of 0.30 ± 0.13 $\mu$m (mean ± std dev), which sets the noise floor of this method and includes image-registration error, vibration and drift between captures. Relative to this floor, the stage performed well as the smallest tested move of 1 $\mu$m was clearly resolved, and repeated executions of the same move agreed to within a few hundred nanometers at every distance, with a standard deviation of about 0.15 $\mu$m for short moves rising to about 0.4 $\mu$m for 250 $\mu$m moves. Because this spread is comparable to the measurement noise, the stage's true repeatability may be better than these values, but the current method cannot show it. While the resolution and repeatability was within the desired specification, the absolute accuracy performed worse. The distance actually traveled typically differed from the commanded distance by a few percent, and the size of this error depended on the direction and location of the move, ranging from roughly 2\% to 9\% for moves of 50 $\mu$m and longer. Because this error is repeatable, it could in principle be reduced by calibration or software correction; although, that has not been tested. Feed rate had little effect on resolution and reversing the direction of travel did not produce a clear backlash signature. Promisingly though, the Z-axis tests produced no X-Y displacement distinguishable from the noise floor, which means our UV exposure should not move relative to our alignment. 

While generally positive, these results have several limitations. At 0.62 $\mu$m per pixel, all reported displacements rely on sub-pixel registration, and no independent displacement reference was used to verify the SIFT model. Error in the pixel-scale calibration, or lens distortion across the field of view could contribute to the measured distance errors, and these have not been separated from errors in the stage kinematics. Because the distance error is repeatable, it could in principle be reduced by a direction- and position-dependent calibration, such as refining the kinematic model parameters; however, no such correction has been implemented or tested.

\subsection*{Vacuum Mounting and Main Frame}

A vacuum mounting platform is added on the end effector to prevent the chip from slipping during motion, shown in Fig. 7. The nanopositioner moves at a maximum 25mm/s, which translates to around 1mN of force dictating the load requirements. The vacuum is generated using a 5V DC pump and is connected to the mounting platform using a thin silicone tube. A pressure-stabilizing chamber is added between the pump and the platform to dampen the pressure waves from the pump, reducing vibration on the lithography platform. The pump's holding force was validated by using a slip-threshold test with a thin wire and calibrated weights. The test was performed with the platform aligned horizontally and vertically to test both load cases. For horizontal loads, the pump maxed out at 58mN while for vertical loads, the maximum was 110mN. Both of these values are well above the expected movement load.

\begin{figure}[h!]
    \centering
    \includegraphics[width=1.0\linewidth]{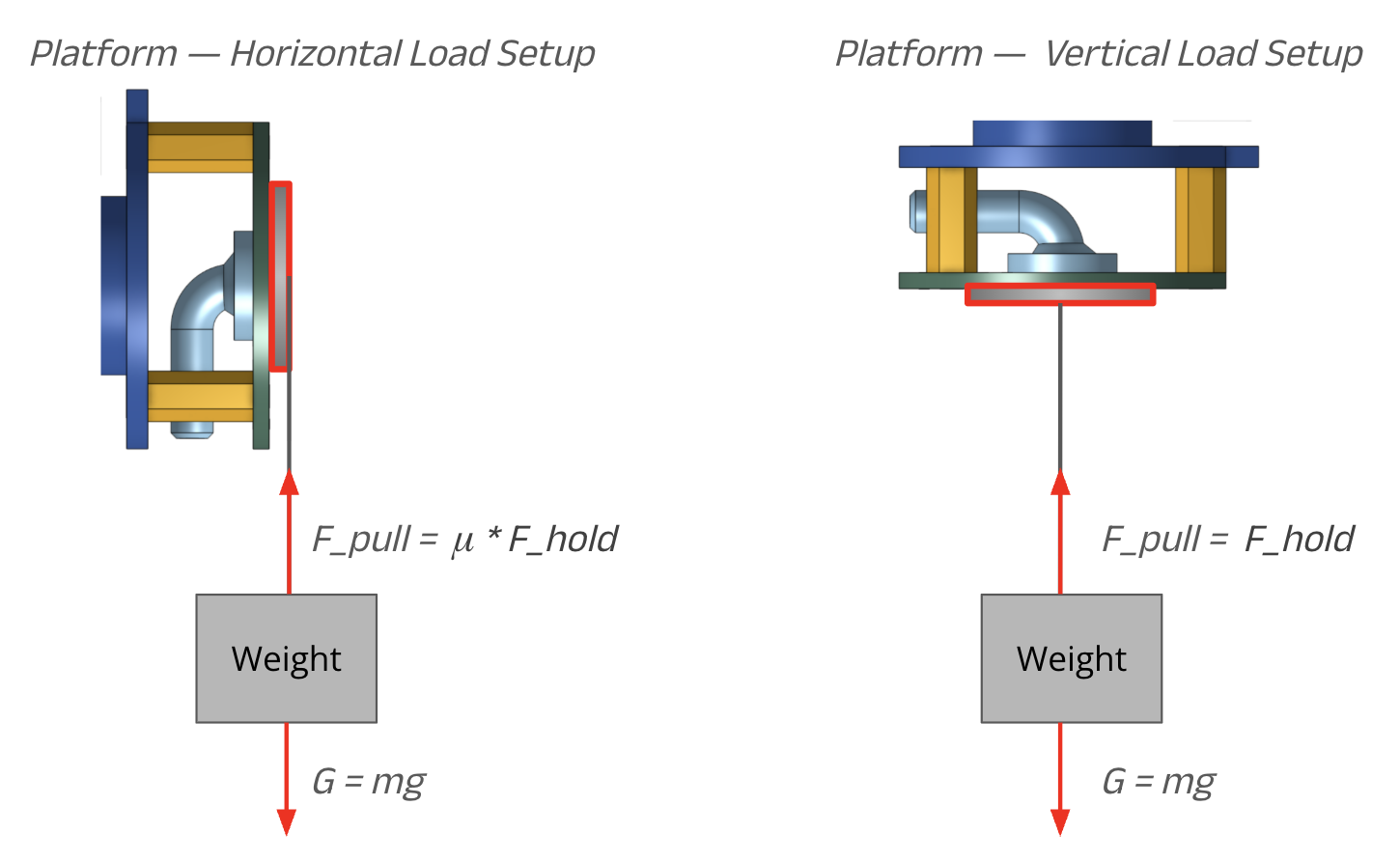}
    \caption{\textbf{Vacuum assembly.} Diagram showing the horizontal and vertical load setups for the vacuum mounting platform.}
    \label{fig:enter-label}
\end{figure}

The full setup is supported by a 3D-printed main frame. The rigidity of the frame is important for reducing vibration and making reliable optical alignment, so a cast or CNC-machined metal frame is the ideal solution. However, that manufacturing process is more complex and less accessible, so a 3D-printable solution was selected instead. The main frame is printed with standard PLA filament, which is widely available, easy to print, and more rigid than alternatives such as PETG or nylon. A triangular infill pattern is used to improve rigidity. For our setup, a 10\% infill density is used; higher infill would make the frame more rigid and reduce vibration, but would increase print time and filament usage.

\subsection*{Software Systems}

Fig. 8 shows our custom GUI software written in Python, which was created to make the patterning experience as easy as possible for the user. The GUI has a live camera view of the chip and stage. To make the stage visible without exposing any photoresist, the user may switch to a red LED light source during positioning. The user may select a series of images to assist with alignment and for near-UV exposure. In red LED mode, the user may preview where their pattern falls on the chip prior to near-UV exposure. Additionally, the GUI integrates controls for the nanopositioner, handling positioning in both absolute coordinates and incremental movement relative to the current position. When the user has aligned the stage and chip to the next pattern, they may switch to near-UV LED mode, which automatically switches the projected pattern to a user-supplied near-UV image (mapped to blue pixels), often containing just fiducial markers for alignment. The user may adjust the Z axis to compensate for differences in the focal distance between red and near-UV light. Once in focus, the user activates exposure, which automatically projects their selected pattern for a duration they specify. After this duration, the projected image is reset to a blank image, ensuring that the photoresist received the right near-UV dosage, and that the chip is ready to be moved to the next position.

\begin{figure}[h!]
    \centering
    \includegraphics[width=1.0\linewidth]{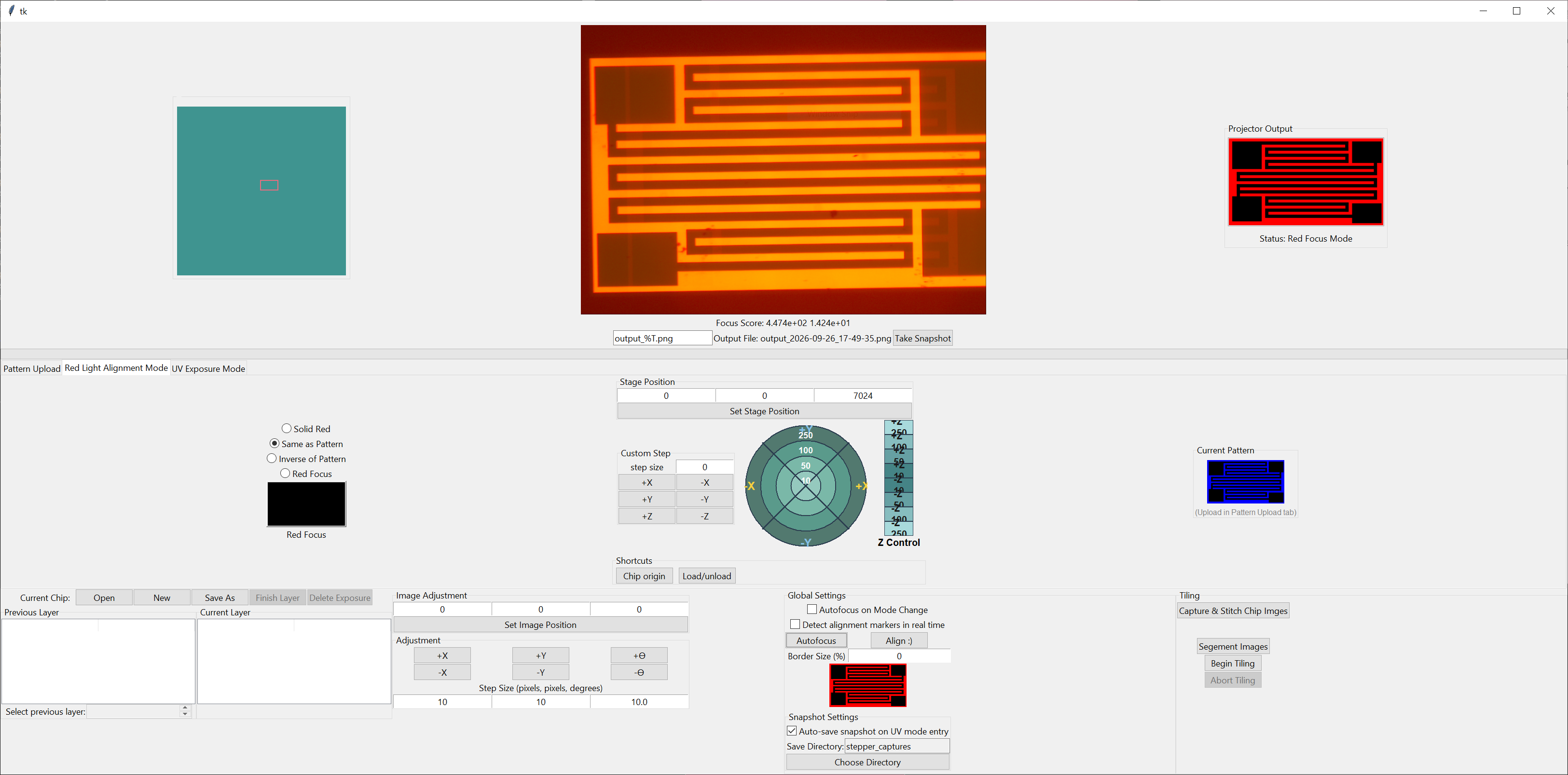}
    \caption{\textbf{Stepper GUI.} A custom user interface was built to allow for easy control of patterning software.}
    \label{fig:enter-label}
\end{figure}

The GUI software also includes useful features such as image transformation, automatic focus, and alignment to on-chip features. Image transformations include rotating, scaling, or translating the projected pattern. The automatic focusing feature samples the camera image at nearby Z-positions and uses a simple Sobel gradient contrast metric to decide which direction sharpens the image; it then takes coarse and fine stage movements until the focus score no longer improves. Lastly, our stepper GUI supports an automated ``stitching and tiling" feature, allowing for larger, more complex circuits to be fabricated more easily as illustrated in Fig. 9.

\begin{figure}[h!]
    \centering
    \includegraphics[width=1.0\linewidth]{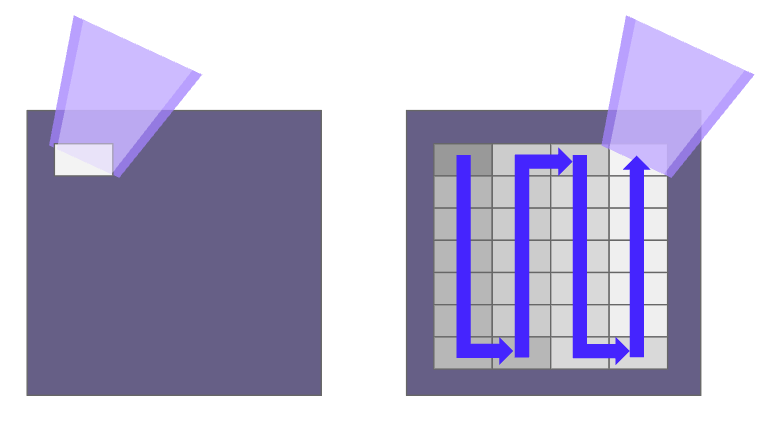}
    \caption{\textbf{Tiling in software.} An illustration of the tiling algorithm to create larger, connected patterns.}
    \label{fig:enter-label}
\end{figure}

To detect the alignment markers on the first layer of a chip, we first note that there is a latent image produced from the exposure of light onto the photoresist; we can pre-process this image using the OpenCV CLAHE algorithm to get better contrast before being fed into the Roboflow RF-DETR model, which locates these markers \cite{10121143} \cite{Zuiderveld1994ContrastLA} \cite{robinson2026rfdetr}. Two simpler approaches were evaluated before adopting a learned detector.

Feature-matching with SIFT and RANSAC was attempted first \cite{lowe2004distinctive} \cite{10.1145/358669.358692}, using dust specs and existing on-chip patterns as keypoints between successive camera snapshots; this failed because repetitive, similarly shaped features at different chip locations produced false-positive matches, and the matcher frequently focused on the featureless outer edge of the camera frame rather than chip content. Template matching (OpenCV matchTemplate against a hand-supplied alignment-marker template \cite{opencv_library}) performed better but remained unreliable, matching only a subset of markers due to variation in hue, intensity, and noise across post-processed images. An object-detection model was adopted to address both failure modes; RF-DETR was selected after evaluation and, trained on a dataset spanning latent, developed, digital, offset, rotated, and tiled marker images, achieved mAP@50 of 89.4\%, precision of 91.7\%, recall of 84.6\%, and F1 of 88.0\%, with 96\%/94\% accuracy on held-out train/test splits. Access to to RF-DETR and its ability to export trained weights was used only once during model training under a trial. The resulting weights were retained permanently, and all inference is performed locally and offline using the open-source onnxruntime Python library, with no ongoing subscription, per-call fee, or dependency on continued access to Roboflow.

For subsequent layers, since the chip will be removed from the stepper tool during process steps, there is a need to perform re-registration between the digital pattern and the pattern that exists on the chip when reinserted. The stage scans an area based on the pattern size before using SIFT and RANSAC to capture and stitch together an image. Then, using the point sets of the stitched image and the digital pattern, a mapping between the point sets is established using the CPD algorithm as shown in Fig. 10, thereby telling us by how much the projected pattern for the next layer must translate and rotate \cite{Myronenko_2010}. We discuss next steps for our software suite in the Discussion section of this paper.

\begin{figure}[h!]
    \centering
    \includegraphics[width=1.0\linewidth]{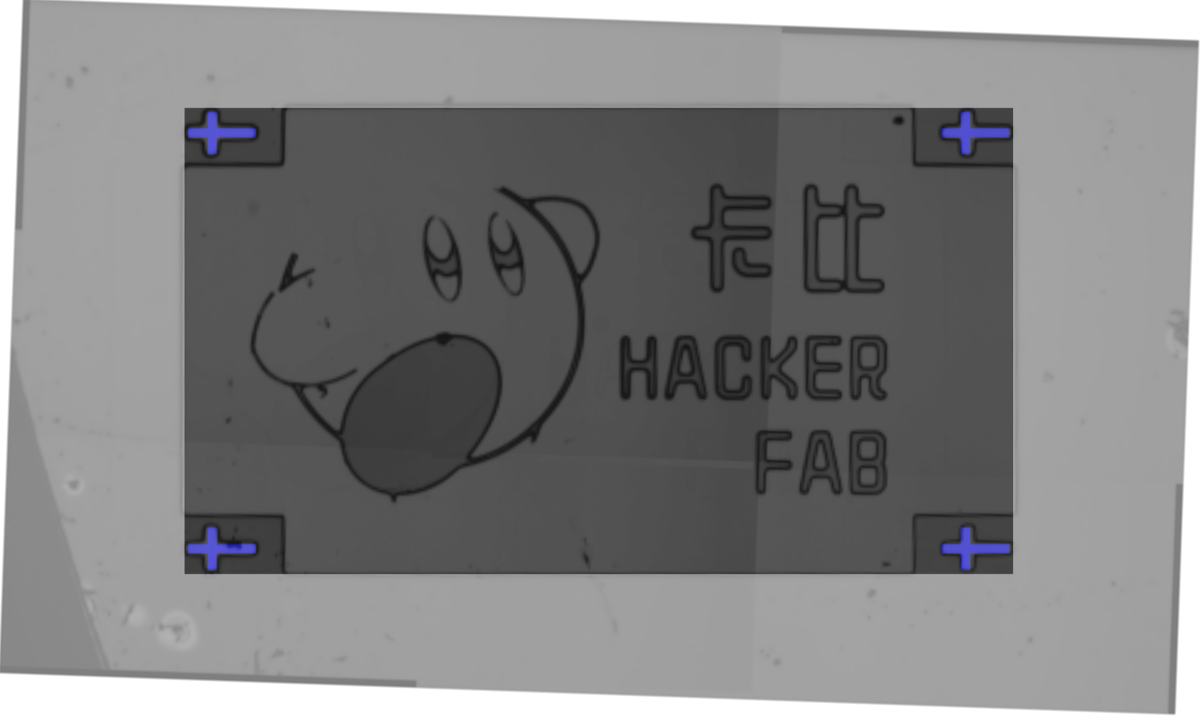}
    \caption{\textbf{Stitched pattern.} Example of aligning a digital pattern of alignment marks onto a stitched image.}
    \label{fig:placeholder}
\end{figure}

\newpage
\section*{Results}

We have characterized the Hacker Fab stepper across several metrics, displayed in Table 1.

\begin{table}[h!]
    \centering
    \begin{tabular}{|c|c|}\hline
Cost& $\sim\$3,000.00$\\\hline
Developed Optical Resolution& 2 $\mu$m\\\hline
Vibration Susceptibility& $\le 0.62 \mu m$\\\hline
Exposure (Die) Dimensions& 1.04mm x 0.58mm\\\hline
Exposure Time& 8 sec\\\hline
Mechanical Step Size& 0.30 ± 0.13 $\mu$m\\\hline
Maximum Sample Size& 1cm x 1cm\\\hline
Tool Dimensions& 36cm x 24cm x 52cm\\\hline
    \end{tabular}
    \caption{\textbf{Stepper Specifications}}
    \label{tab:my_label}
\end{table}

According to the Rayleigh criterion shown in Equation 1, our patterning wavelength ($\lambda$) of 410nm and numerical aperture (NA) of 0.25 give a theoretical critical dimension (CD) of 1 $\mu$m.

\begin{equation}
    CD=0.61\frac{\lambda}{NA}
\end{equation}

We have achieved a practical optical resolution of 2 $\mu$m, as shown in the test patterns in Figs. 11 and 12, using AZ P4110 photoresist and AZ 400K developer. The stepper also has an alignment accuracy of 5 $\mu$m at a cost of \$3000. For comparison, a Heidelberg $\mu$MLA maskless aligner can achieve minimum feature sizes down to 1 $\mu$m and has a mechanical alignment of 0.5 $\mu$m, but at a minimum cost of \$100,000 \cite{umla2025}. A Karl Suss MA6 contact aligner uses a synthesized physical mask to transfer a pattern to the substrate; while the tool is capable of achieving a patterning resolution similar to our stepper (1 $\mu$m) and an alignment accuracy of 0.5 $\mu$m, these masks are expensive and time-consuming to generate, and our stepper can create patterns within seconds of exposure \cite{ma62025}. As considered in the Discussion section, our stepper can be improved to target sub-micron resolution and alignment at a fraction of the cost of commercially available products.

\begin{figure}[h!]
    \centering
    \includegraphics[width=1.0\linewidth]{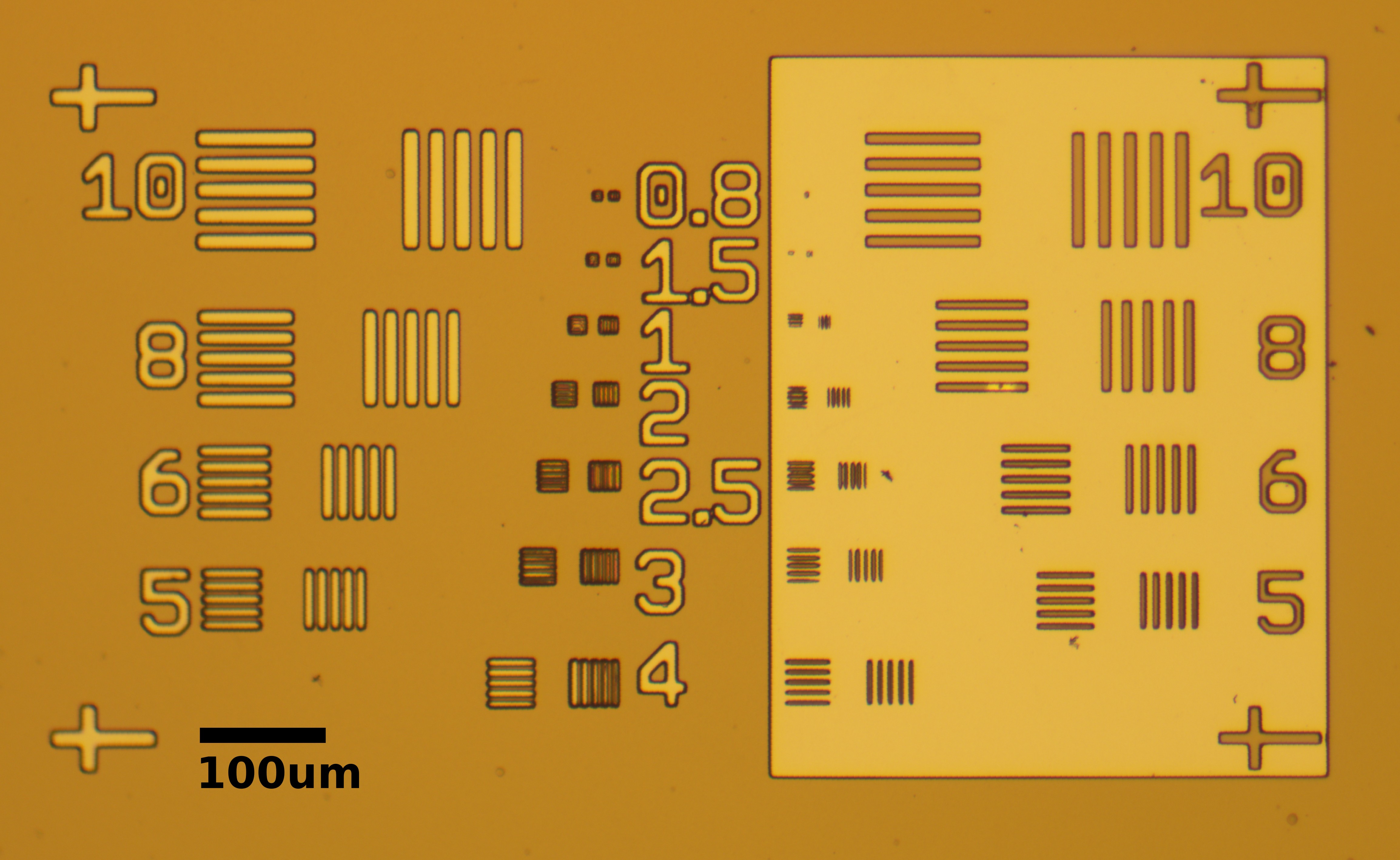}
    \caption{\textbf{Test Pattern, Example 1.} Several test structures patterned and measured using our photolithography stepper and camera software.}
    \label{fig:enter-label}
\end{figure}

\begin{figure}[h!]
    \centering
    \includegraphics[width=1.0\linewidth]{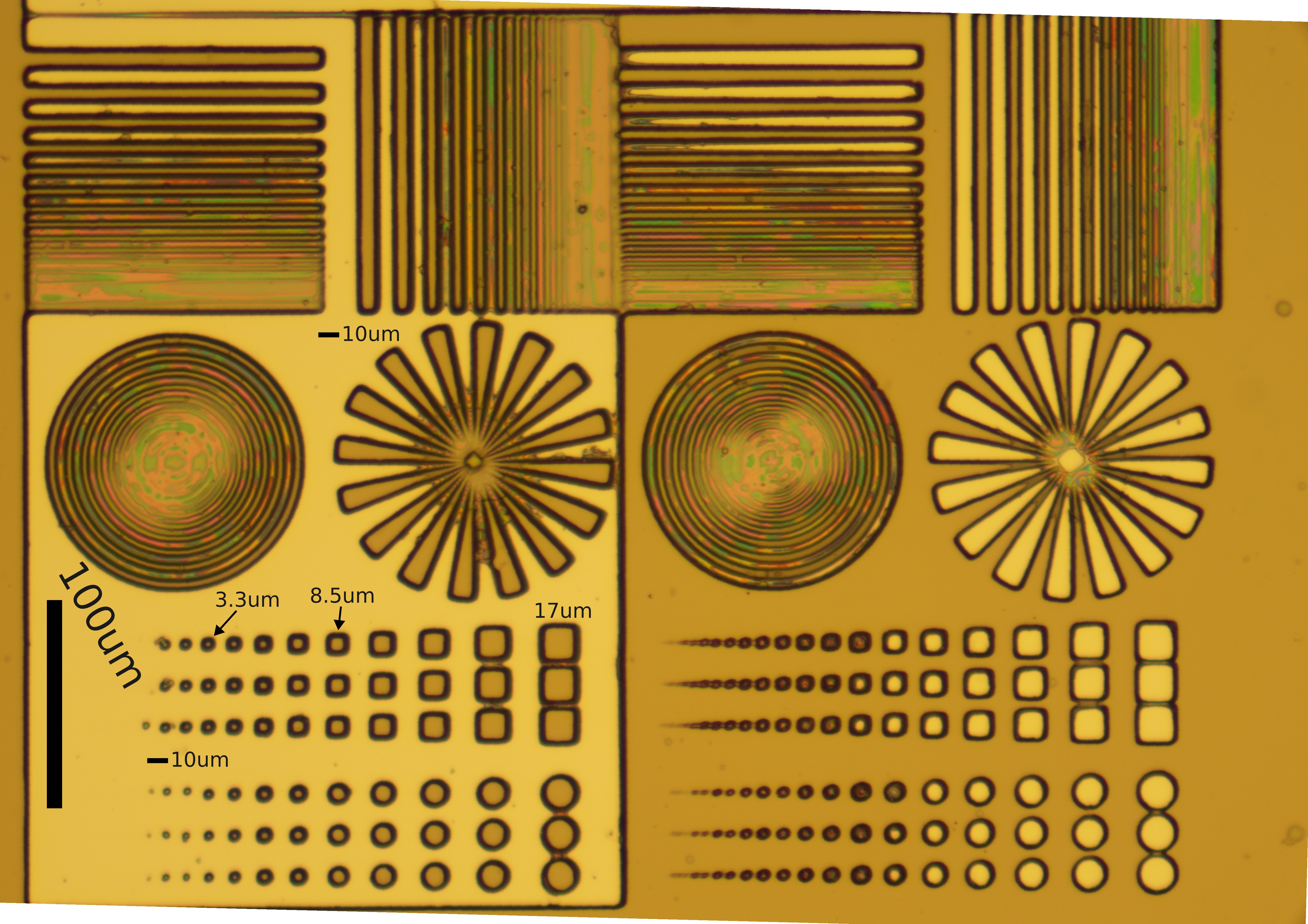}
    \caption{\textbf{Test Pattern, Example 2.} Several test structures patterned and measured using our photolithography stepper and camera software.}
    \label{fig:enter-label}
\end{figure}

We have demonstrated the ability of our stepper to be used for semiconductor process flows. In the Hacker Fab at Carnegie Mellon University, the stepper has been used as part of a photolithography process to reliably create NMOS transistors with a gate length as small as 5 $\mu$m. The stepper is used at multiple points in the NMOS process, including patterning of the gate, source, and drain regions, along with patterning of aluminum contact pads (Fig. 13, 14, 15) \cite{nmos2024}.

\begin{figure}[h!]
    \centering
    \includegraphics[width=1.0\linewidth]{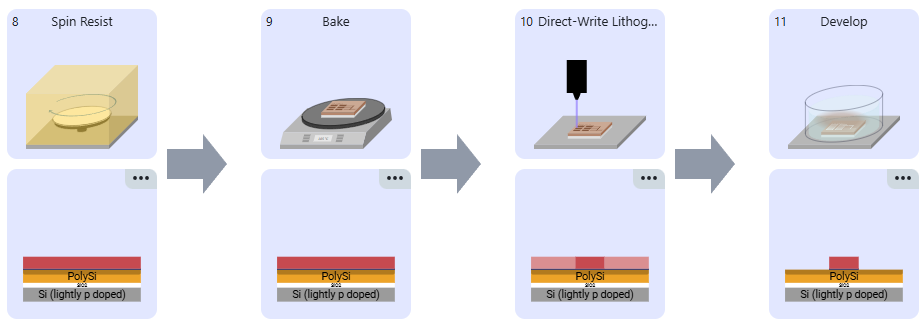}
    \caption{\textbf{Gate Patterning.}}
    \label{fig:enter-label}
\end{figure}

\begin{figure}[h!]
    \centering
    \includegraphics[width=1.0\linewidth]{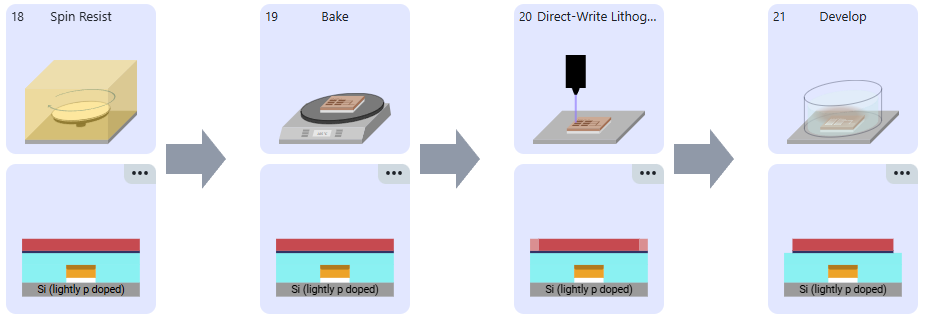}
    \caption{\textbf{Source/Drain Patterning.}}
    \label{fig:enter-label}
\end{figure}

\begin{figure}[h!]
    \centering
    \includegraphics[width=1.0\linewidth]{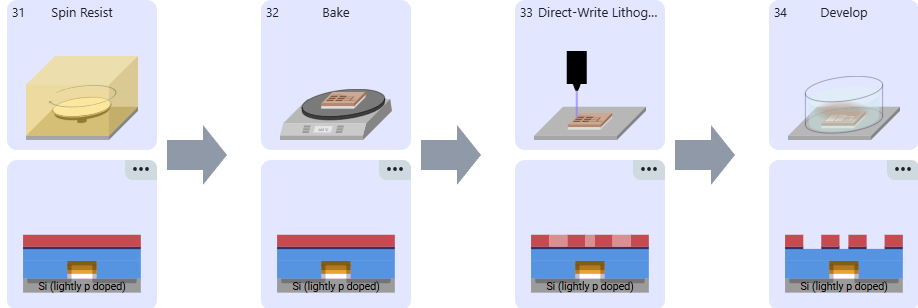}
    \caption{\textbf{Metal Contact Patterning.}}
    \label{fig:enter-label}
\end{figure}

A set of fabricated NMOS transistors are shown alongside a measured IV curve in Figs. 16, 17, and 18. In each of these process steps, the stepper is used to create a photoresist pattern on top of and aligned to underlying features. The patterns are first generated by spin-coating and baking a layer of AZ P4210 photoresist followed by alignment and exposure. The resulting images in the photoresist are developed by submerging the chip in AZ400K developer solution, and the patterns are inspected optically before proceeding to the next step.

\begin{figure}[h!]
    \centering
    \includegraphics[width=1.0\linewidth]{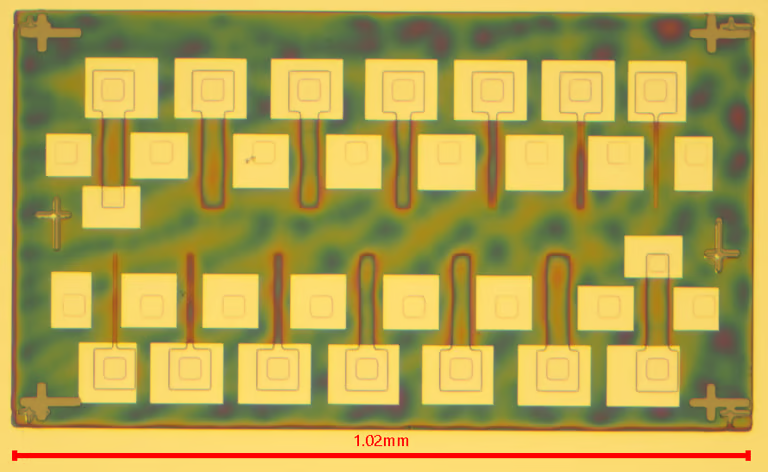}
    \caption{\textbf{Patterned NMOS devices.} A measured gate length of 5 microns is visible in these NMOS devices.}
    \label{fig:enter-label}
\end{figure}

\begin{figure}[h!]
    \centering
    \includegraphics[width=1.0\linewidth]{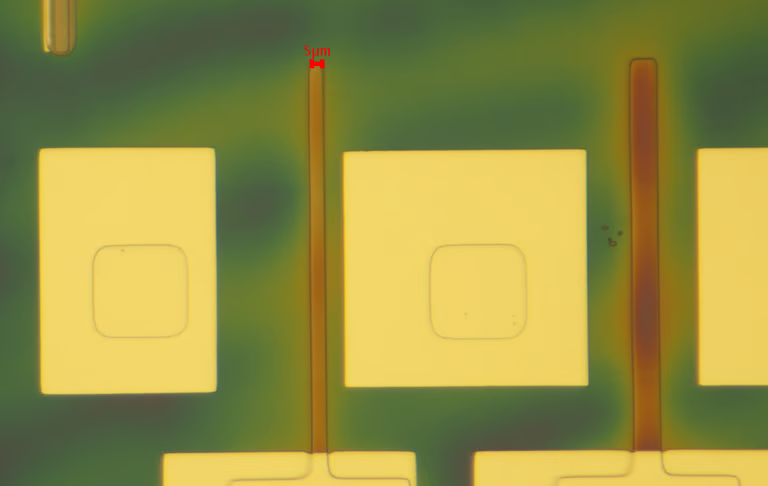}
    \caption{\textbf{Measurement of gate length showing 5 microns.}}
    \label{fig:enter-label}
\end{figure}

\begin{figure}[h!]
    \centering
    \includegraphics[width=1.0\linewidth]{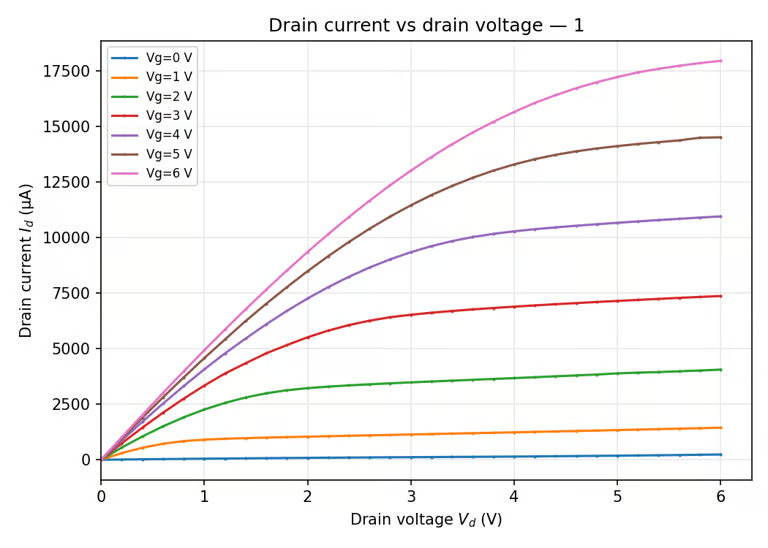}
    \caption{\textbf{IV curves from NMOS devices fabricated using the Hacker Fab stepper.}}
    \label{fig:enter-label}
\end{figure}

The stepper is an integral part of the NMOS process in the Hacker Fab, and has proven to be a reliable tool for patterning. The stepper has been used to create hundreds of devices at the Carnegie Mellon Hacker Fab, and the build has been successfully replicated at other universities such as the University of Waterloo \cite{uw2026} and the University of California, Irvine \cite{uci2026}. This is achieved thanks in large part to the low-cost of the build, as well as the open-source nature of the project.

\newpage
\section*{Discussion}
As demonstrated in this paper, the potential for open-source hardware is monumental. Through initiatives such as the Hacker Fab at Carnegie Mellon, open-source hardware allows engineers to collaborate and exchange ideas with each other to develop the best solutions to problems.

There are still many ways our stepper can be improved. One such approach would be to modify the optical path of the machine. Instead of using a TI evaluation board, which contains various peripherals such as HDMI that are unnecessary for our applications, a custom PCB can be built using the TI chipset. The DLP301S / DLPC1438 chipset could be used as it is designed for resin 3D printers, appropriate for use with 405nm near-UV light. It can also receive a ``parallel video" protocol, which is simpler to generate for processing raw video signals than the HDMI protocol that the 471TP uses. It is estimated that the cost of the projector subsystem could decrease from \$1000 to less than \$650 by making a custom PCB \cite{gitbook2025}. Water immersion photolithography and optical proximity correction may also be explored as means to achieve higher resolution in patterning, though there is a trade-off between maximum reticle size and minimum feature size in the former technique.

Another approach to improve the stepper would be the refinement of the positioner stage. As described in the Materials and Methods section, the nanopositioner is a very robust system for our application of photolithography. However, more work remains in characterization of the machine, including validating the measurements against a laser interferometer, repeating them at higher magnification, randomizing the order of feed rates, and implementing and verifying a kinematic calibration. To this end, a low-cost interferometer by the same developer is in progress, allowing for nanometer measurement of displacement which can be used for feedback control \cite{interferometer2026}. Our team plans to continue work on this subproject for more reliable and accurate positioning.

Lastly, for the GUI, we wish to finalize the implementation of tiling to enable full automation of the patterning process. What remains is to complete multilayer integration of the tiling feature, which would allow users to automatically align to previous layers of their chip via the GUI. The aforementioned nanopositioner system provides more careful alignment to avoid overlapping exposures in the tiling sequence; the addition of a rotational stage also allows for better correction of rotational errors (as opposed to relying on rotation of the digital image projection), while also creating flexibility for stage movement capabilities. In addition, improvements can be made to the movement calculations in detecting fiducial markers to make the system more robust. Our team intends to continue work on the software suite to provide a fast, seamless patterning experience for users of the stepper.

Lastly, our photolithography system can be applied to countless fields beyond traditional NMOS devices. While the Hacker Fab continues to use the stepper to build NMOS and eventually CMOS devices, our tool may easily be applied to other research areas in microelectromechanical systems and microfluidics as well as photonics. The ability of our stepper tool to pattern single-digit micron length features in seconds, without the need for a physical photomask, makes it an excellent tool for rapid prototyping of MEMS and microfluidic applications; although we have only experimented with AZ P4210 photoresist in the Hacker Fab, other photoresists such as SU-8 can be paired with an appropriate-wavelength LED in our system to create the desired structures for MEMS and microfluidics \cite{10.1021/acsomega.3c05544}. Likewise, our photolithography system can be applied to the field of photonics, allowing for quick fabrication of common structures such as waveguides, gratings, and resonators \cite{Zheng2021}. Thus, we aspire for our tool to be a valuable addition to any laboratory that wishes to explore the rapid prototyping of devices at the micro- and nanoscale.

\section*{Conclusion}

Our contribution offers a maskless photolithography stepper that can be built for \$3000, which is capable of aligning and patterning down to the single-digit micron-scale. Further improvements to the stepper's optics, electromechanical design, and software systems can push the tool's patterning and positioning resolution to the sub-micron regime. We hope that our emphasis on an open-source solution to microfabrication can help lower the barrier to entry for micro- and nanofabrication, thereby revolutionizing the semiconductor industry.

\section*{Acknowledgments}
The authors acknowledge Sam Zeloof, who prototyped initial concepts of the stepper design, as well as Alexander Hakim, Jay Kunselman, and Prof. Tathagata Srimani of Carnegie Mellon University, whose efforts were instrumental in the development of the Hacker Fab. In addition, the authors acknowledge Carson Swoveland, Sky Bailey, Luca Garlati, Italia Grant, Kashvi Verma, Jiangtian Luo, and Bo Ying Su of Carnegie Mellon University for their work on the stepper project over the past several years. Lastly, we thank John Peterson of the University of Utah for his contributions to the project's documentation and photos.

\section*{Author Contributions}

\begin{itemize}
    \item \textbf{Conceptualization:} Elio Bourcart, B. Joel Gonzalez, J. Kent Wirant, Matthew T. Moneck
    \item \textbf{Data Curation:} J. Kent Wirant, Rachel Jiang, Peiyu Lin, Michael Casey Honaker, Toby Yang, Michael Juan, Justin Wang
    \item \textbf{Formal Analysis:} J. Kent Wirant, Rachel Jiang, Peiyu Lin, Michael Casey Honaker, Toby Yang, Michael Juan, Justin Wang, William Gao
    \item \textbf{Funding Acquisition:} Elio Bourcart, Matthew T. Moneck
    \item \textbf{Investigation:} Elio Bourcart, B. Joel Gonzalez, J. Kent Wirant, Rachel Jiang, Peiyu Lin, Michael Casey Honaker, Toby Yang, Michael Juan, Justin Wang, William Gao
    \item \textbf{Methodology}: Elio Bourcart, B. Joel Gonzalez, J. Kent Wirant, Rachel Jiang, Peiyu Lin, Michael Casey Honaker, Toby Yang, Michael Juan, Justin Wang, William Gao
    \item \textbf{Project Administration:} B. Joel Gonzalez, Elio Bourcart, J. Kent Wirant, Matthew T. Moneck
    \item \textbf{Resources:} B. Joel Gonzalez, Elio Bourcart, Matthew T. Moneck
    \item \textbf{Software:} J. Kent Wirant, Rachel Jiang, Peiyu Lin, Jessica Alacce, Jocelyn Xu, Justin Wang, Willaim Gao, B. Joel Gonzalez
    \item \textbf{Supervision:} B. Joel Gonzalez, Elio Bourcart, J. Kent Wirant, Matthew T. Moneck
    \item \textbf{Validation:} Elio Bourcart, J. Kent Wirant, Rachel Jiang, Peiyu Lin, Michael Casey Honaker, Toby Yang, Michael Juan, Justin Wang
    \item \textbf{Visualization:} J. Kent Wirant, Rachel Jiang, Peiyu Lin, Jessica Alacce, Jocelyn Xu, Justin Wang
    \item \textbf{Writing - Original Draft:} B. Joel Gonzalez
    \item \textbf{Writing - Review and Editing:} J. Kent Wirant, B. Joel Gonzalez, Michael Casey Honaker, Rachel Jiang, Matthew T. Moneck
\end{itemize}


\bibliography{plos_one_hfab_stepper.bib}

\end{document}